\documentclass{article}

\usepackage[nonatbib, preprint]{neurips_2020}

\usepackage[utf8]{inputenc} 
\usepackage[T1]{fontenc}    
\usepackage{hyperref}       
\usepackage{url}            
\usepackage{booktabs}       
\usepackage{amsfonts}       
\usepackage{nicefrac}       
\usepackage{microtype}      
\usepackage{graphicx}       
\usepackage{color, colortbl} 
\usepackage{multirow} 
\usepackage{longtable}

\title{Target Privacy Threat Modeling for COVID-19 Exposure Notification Systems}

\author{
  Ananya Gangavarapu\\
  PathCheck Foundation\\
  Cambridge, MA, USA\\
  \texttt{ananya.gangavarapu@pathcheck.org} \\
 \And
  Ellie Daw \\
 PathCheck Foundation\\
  Cambridge, MA, USA\\
  \texttt{elliemdaw@gmail.com} \\
\And
 Abhishek Singh\\
 PathCheck Foundation\\
 Cambridge, MA, USA\\
  \texttt{abhishek.singh@pathcheck.org} \\
 \And
 Rohan Iyer \\
 PathCheck Foundation\\
 Cambridge, MA, USA\\
  \texttt{rohan.iyer@pathcheck.org} \\
 \And
  Gabriel Harp\\
  The MIT Press\\
  Cambridge, MA, USA\\
  \texttt{gabriel.harp@pathcheck.org} \\
 \And
  Sam Zimmerman\\
 PathCheck Foundation\\
  Cambridge, MA, USA\\
  \texttt{sam@pathcheck.org} \\
\And
  Ramesh Raskar\\
 PathCheck Foundation\\
  Cambridge, MA, USA\\
  \texttt{ramesh.raskar@pathcheck.org} \\
}

\begin{document}

\maketitle

\begin{abstract}

The adoption of digital contact tracing (DCT) technology during the COVID-19 pandemic has shown multiple benefits, including helping to slow the spread of an infectious disease and to improve the dissemination of accurate information. However, to support both ethical technology deployment and user adoption, privacy must be at the forefront. With the loss of privacy being a critical threat, thorough threat modeling will help us to strategize and protect privacy as digital contact tracing technologies advance.

Various threat modeling frameworks exist today, such as LINDDUN, STRIDE, PASTA, and NIST, which focus on software system privacy, system security, application security, and data-centric risk, respectively. When applied to the exposure notification system (ENS) context, these models provide a thorough view of the software side but fall short in addressing the integrated nature of hardware, humans, regulations, and software involved in such systems. Our approach addresses ENSs as a whole and provides a model that addresses the privacy complexities of a multi-faceted solution. We define privacy principles, privacy threats, attacker capabilities, and a comprehensive threat model. Finally, we outline threat mitigation strategies that address the various threats defined in our model.

\end{abstract}

\section{Introduction}

Privacy is one of the most important components when evaluating a digital contact tracing system. Privacy is of paramount importance due to multiple reasons ranging from government policies and regulations to adoption by individuals. Existing threat modeling frameworks and approaches focus mostly on security threats and effectiveness during the software development life-cycle (SDLC). Digital tracking of COVID-19 infections requires leveraging the existing components, such as web server technologies, to build new services such as exposure notification (EN) apps. The repurposing of existing components and the algorithmic nature of DCT leads to radically different privacy concerns not considered while designing or using those components in traditional ways. Hence, an enhanced approach is needed for privacy threat modeling and privacy impact assessment for CT capabilities.

We have started with evaluating existing threat modeling frameworks like LINDDUN \cite{LINDDUN}, STRIDE \cite{STRIDE}, PASTA \cite{PASTA}, and NIST \cite{NIST} for use in privacy threat modeling. Most of the privacy threat models provide a software-centric approach to threat modeling and fall short when it comes to non-stereotypical privacy attacks like coercion attacks or complying with evolving privacy directives such as ePrivacy directives from European Data Privacy Board  (EDPB) \cite{eprivacy}.  To address the additional privacy requirements, we propose a few enhancements to existing privacy threat models. First, we propose a modeling approach based on target state attacker capabilities and applicable exhaustive threat categories. The target state ensures the minimization of false positives, the minimization of overlooked threats, and a consistent result regardless of who is doing the threat modeling. Another proposed enhancement is to simplify threat models to accommodate many-to-many relationships between threat categories and attacker capabilities. Finally, in order to avoid confusion, we have standardized the terms used across target privacy threat models. 

This document first introduces the enhanced framework and then uses the framework to build the Target Privacy Threat Model for Exposure Notification System (ENS).

\section{Terminology}

There are a variety of ways to leverage technology to aid the manual contact tracing process. Traditionally, this process involves a public health official interviewing a person who has been diagnosed with an infectious disease in order to identify who they may have had contact with so that these contacts can be alerted to take proper measures to prevent further spread of the virus. Apps that have been built to help with this process typically alert users who have had a potential virus exposure, rather than performing tracing of contacts. As such, these apps are often referred to as exposure notification (EN) or exposure alerting (EA) apps. The functionality provided works in conjunction with manual contact tracing to scale the efforts of early quarantine and other preventive measures. Some applications provide additional features beyond exposure notification and therefore may not be pure EN apps. In this paper, we use the umbrella term exposure notification systems (ENS) to refer to EN, EA, and apps with functionality to aid in scaling outcomes of manual contact tracing, though it is important to note that privacy-preserving applications under this umbrella do not typically “trace” users and their contacts.

\section{Target Privacy Threat Modeling Framework}

The Target Privacy Threat Modeling (TPTM) Framework is a privacy-driven threat modeling framework that helps to systematically identify privacy attacker capabilities and mitigate privacy threats in a system or an interconnected platform. The TPTM Framework leverages an attacker-centric threat model that is driven by the attacker’s capabilities and motivations to exploit the privacy vulnerabilities. Also, as a privacy threat model framework, the focus is on attacker goals that compromise the privacy of users, rather than their security. 

The following diagram shows various steps of the TPTM Framework:

\begin{figure}[htp]
    \centering
    \includegraphics[scale=0.2]{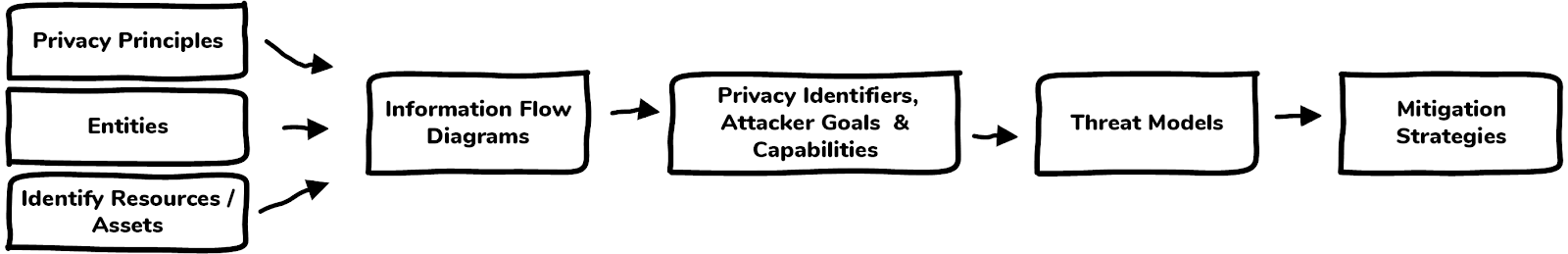}
    \label{fig:graphic1}
\end{figure}

The TPTM Framework starts with privacy principles and identifies the resources and/or assets in the scope. Threat models help to identify attacker goals and capabilities in the context of specific information flow diagrams and develop effective mitigation strategies. The TPTM Framework recommends doing threat modeling at various phases of the development and deployment of the systems.

\section{Target Privacy Threat Modeling Framework for Exposure Notification Systems}

ENS, which has become an important toolset in mitigating COVID-19, has two principal components:
\begin{itemize}
    \item \textbf{Case management:} Key functions are i) streamlined electronic capture and management of data on patient and contacts and ii) integrated workflows with surveillance systems.
    \item \textbf{Proximity tracing / exposure notification:} Key functions are i) use opt-in digital tools to augment manual contact tracing and ii) use Bluetooth technologies to compute proximity and duration of exposure to patients diagnosed with COVID-19.
\end{itemize}

\subsection{ENS Architectural Approaches}

All of the ENS solution implementations fall into one of three categories of architecture:
\begin{itemize}
    \item \textbf{A centralized architecture:} The anonymity of users and confidentiality of contact events is only provided with regard to outside entities, i.e., other users or external actors; the operators and involved authorities can, however, identify all users and connect them to recorded contact histories.
    \item \textbf{A partially decentralized architecture:} Users and contact events are only concealed from other users and third parties; the server can de-anonymize positively tested users. The system also has a data donation function, by which users can choose to share their contact histories for epidemiological research. If done so, positively tested users’ contact events would become visible to operators and authorities.
    \item \textbf{A completely decentralized architecture:} Users remain anonymous toward other users and third parties, and their contact events also remain private. Operators and authorities can de-anonymize positively tested users but cannot access their contact history. Epidemiological research is not possible.
\end{itemize}

Each of the architectures mentioned above results in different threat models and hence different mitigation strategies. Before we get into specifics of threat models, let us first identify privacy principles. 

\subsection{Privacy Principles}

Privacy principles form the base for the scope and key business requirements of ENS approaches. General Data Protection Regulation (GDPR) has been the gold standard for personal data protection, and this document uses GDPR and related privacy directives from EDPB. The following are key privacy principles used as the scope for the target threat modeling:

\begin{itemize}
    \item \textbf{Lawful:} ENS and related components must comply with all applicable laws, rules, and regulations. Any data collection and use must have a lawful basis.
    \item \textbf{Informed consent:} A user must provide informed consent as a prerequisite for the installation and use of ENS. A user should be able to give their consent to each function of an app separately, such as the collection of proximity data, location data, sharing data, or other key separate functions.
    \item \textbf{Purpose binding:} Purpose binding ensures that personal data processing is performed according to predetermined purposes.
    \item \textbf{Identity control:} Users make the determination to release redacted, disconnected, and/or aggregated space-time points from location data, or obfuscated identifiers from e.g., Bluetooth.
    \item \textbf{Transparency:} Consumers should be given notice of an organization’s information practices before any personal information is collected from them.
    \item \textbf{Accountability:} In order to ensure that organizations adhere to ENS privacy principles, there must be enforcement measures.
    \item \textbf{Proportionality:} The potential risk that private information may be exposed to or misused as a result of the ENS must be proportional to the public health benefits of that system for combating the epidemic.
    \item \textbf{Data retention:} To minimize risk of improper data use, breach, or loss, ENS must have responsible data retention procedures, e.g., practicing data minimization and destroying data after a set amount of time.
\end{itemize}

\subsection{Entities}

The main entities for privacy threat modeling are the data subject, the data controller, and the data processor(s). The following table depicts different actors under each of the entity groups:

\definecolor{Gray}{rgb}{0.8, 0.8, 0.8}

\begin{center}
\begin{tabular}{|p{0.2\textwidth}|p{0.7\textwidth}|}
 \hline
 \textbf{Entity Type} \cellcolor{Gray}
 & \textbf{Actors and Participants} \cellcolor{Gray}
 \\ 
 \hline
 Data Subject & Users or business entities of ENS or related resources
 \\ 
 \hline
 Controller &
 Health authorities, government, and other parties determining purposes and means of processing personal data
 \\
 \hline
 Processor &
 Organizations providing resources for data processing, such as cloud computing providers, database providers, app providers, etc.
 \\
 \hline
\end{tabular}
\end{center}

\subsection{Identify resources and assets}

Resources are typically software components or tools used in ENS. In addition to privacy principles, the assets and the resources drive the scope of the target threat modeling activities. The following table lists all of the components/services used in each of the architectural approaches. 

\begin{center}
\begin{tabular}{|p{0.3\textwidth}|p{0.4\textwidth}|p{0.3\textwidth}|}
 \hline
 \textbf{Architectural Approach} \cellcolor{Gray}
 & \textbf{Related Resources/Assets} \cellcolor{Gray}
 & \textbf{Example} \cellcolor{Gray}
 \\ 
 \hline
 
Centralized Architecture &
Web server, browser, mobile app, databases, algorithms, and app stores. &
Singapore’s TraceTogether \cite{opentrace}

 \\
 \hline
 
Partially Decentralized Architecture &
Web server, browser, mobile app, databases, algorithms, app stores, and support services. &
PathCheck Foundation apps \cite{pathcheck}, Germany’s Corona Warn \cite{coronawarn}, Ireland’s Covid Tracker \cite{HSEIreland}

 \\
 \hline
 
Decentralized Architecture &
Mobile app, app stores, and support services. &
TraceCorona \cite{tracecorona}
\\
\hline
\end{tabular}
\end{center}

Resources or assets are evolving as time passes and all the vendors, including Google and Apple, are updating the software and other assets continuously. For this paper, we are considering the following versions / frameworks only:
\begin{itemize}
    \item Google Exposure Notification Server v0.7.0
    \item GAEN Exposure Notification API v1.5 \cite{leith}
    \item Pan-European Consortium (PEPP-PT)
    \item DP-3T
    \item Apple iOS 11 or greater
    \item Android 8 or greater
\end{itemize}

For target threat models, we take superset of the resources required for all three architectural approaches. 

\subsection{Information Flow Diagrams (IFDs)}

Data flow diagrams (DFDs) traditionally used to identify privacy threats are not effective for privacy identifiers in ENS. DFD-based threat modeling fundamentally looks at how data is designed to move through a system. The approach cannot, therefore, provide a means to inherently analyze how an application appears to a potential attacker. Instead, we propose information flow diagrams, IFDs, to identify all privacy identifiers in functional and/or business terms. 

The figure below depicts an IFD for a partially decentralized architecture (GAEN mobile app / Express Server):

\begin{figure}[htp]
    \centering
    \includegraphics[scale=0.2]{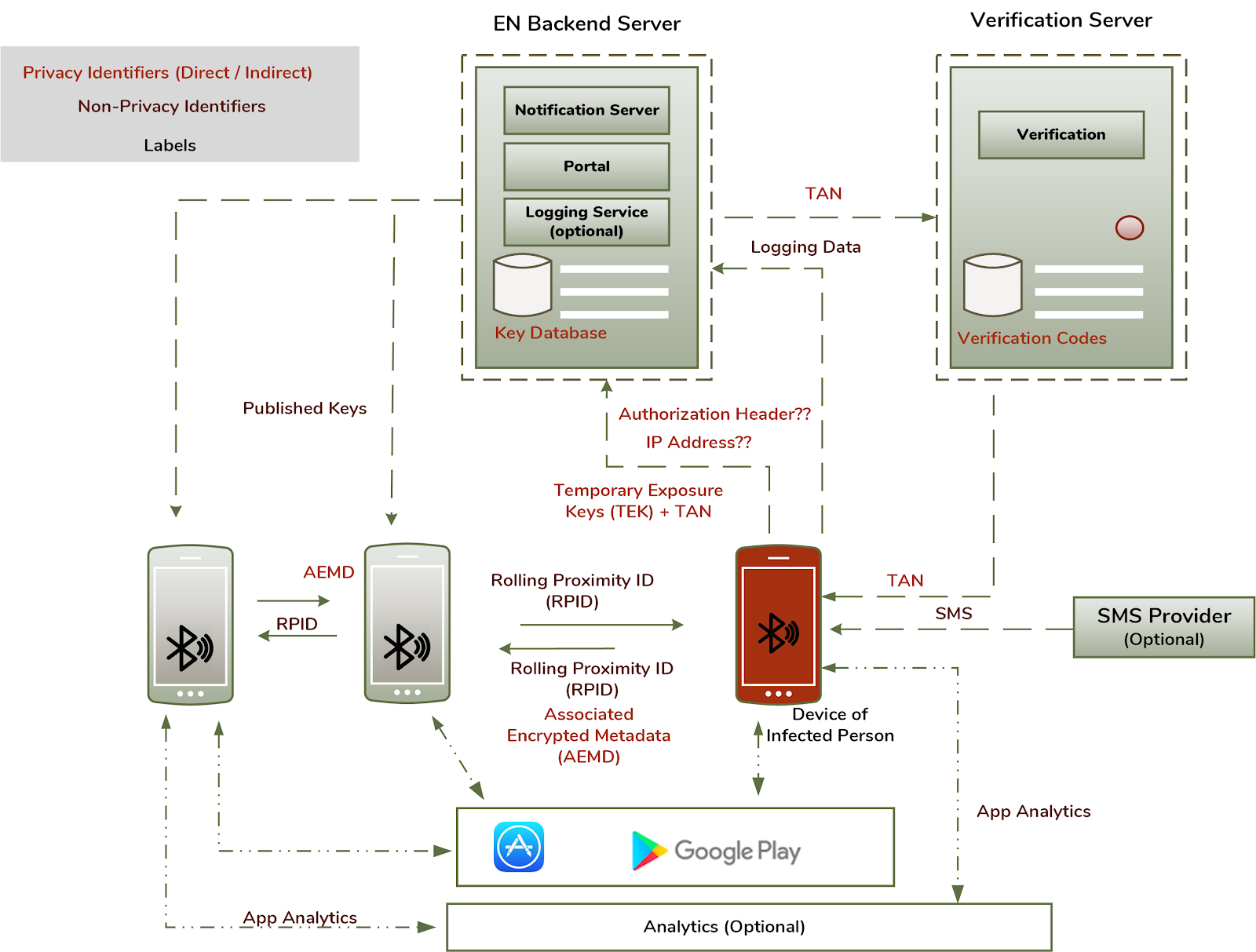}
    \label{fig:graphic2}
\end{figure}

\subsection{Privacy Identifiers}
Privacy Identifiers are any data, element, or information that can be used, directly or as a proxy,  to identify a person or group of persons.  The following table lists most of the direct and indirect privacy identifiers in the target state of ENS.

\begin{center}
\begin{tabular}{|p{0.5\textwidth}|p{0.5\textwidth}|}
 \hline
 \textbf{Direct Privacy Identifiers }\cellcolor{Gray}
 & \textbf{Indirect Privacy Identifiers} \cellcolor{Gray}
 \\ 
 \hline
 Name, address, user email, phone IMEI, IP address, MAC address
 &
 Location data, local public health agency, age of the user, infection status, device model / version, gender, date of interview, temporary encryption key (TEK), associated encrypted metadata (AEMD), pseudonyms
 \\
 \hline
\end{tabular}
\end{center}

Several non-identifiers are being used in ENS as well, such as transaction logging data, OS versions, Locale, TAN, etc.

\subsection{Privacy Threat Categories}

Privacy threat categories provide high-level goals of the attacker(s). In traditional privacy threat modeling approaches, threat categories are identified separately from the attacker's goals and we combine both of them for easier reference.

\begin{itemize}
    \item \textbf{Surveillance} is the observation or monitoring of an individual’s communications or activities. 
    \item \textbf{Stored data compromise} refers to end systems that do not take adequate measures to secure stored data from unauthorized or inappropriate access.
    \item \textbf{Misattribution} occurs when data or communications related to one individual are attributed to another.
    \item \textbf{Secondary use} is the use of collected information about an individual without the individual’s consent for a purpose different from that for which the information was collected.
    \item \textbf{Exclusion} is the failure to allow individuals to know about the data that others have about them or to participate in its handling and use.
    \item \textbf{Linkability} is a higher-level goal of a privacy attacker and refers to the ability to identify related items in two (or more) seemingly unrelated datasets. From an attacker's perspective, this is done with the intention of identifying a person or a location. For example: identifying a place of worship based on the times when groups of people gathered there.
    \item \textbf{Identification} is a higher-level goal of a privacy attacker. It uses adversarial tactics to identify or de-anonymize a person or a group of people. This can either be done with the data directly, or by some combination or observation of various types of datasets. It can also occur during any phase of the information flow (or app usage in this case).
    \item \textbf{Detection} is when an attacker is able to detect whether a piece of data exists or not. A common example of this is membership inference, where the attacker is able to tell whether or not a particular individual is present within a dataset.
    \item \textbf{Non-repudiation} occurs when a user is unable to refute that a piece of data belongs to them. In the context of digital exposure notification, imagine a user being unable to deny that a certain anonymous value was transmitted by their device.
    \item \textbf{Integrity compromise} is a higher-level goal of a privacy attacker, as well. These attacks aim to compromise the validity or trustworthiness of infection or location data. This can happen during the broadcast phase, when an attacker may aim to broadcast dishonest values or broadcast values with dishonest metadata/properties. It could happen during the reporting phase, when an attacker may aim to report a dishonest positive infection result. It could also happen during the download phase, when an attacker may aim to block downloads of honest values.
\end{itemize}

\subsection{Attacker Capabilities}

These are means by which some attackers may be able to achieve attacker goals. The attacker capabilities are not completely mutually exclusive to each other and may result in overlaps in threat models.

\textbf{Coercion attacks} are especially threatening during COVID-19. Coercion is a unique type of privacy threat that aims to manipulate a user in order to divulge information that compromises privacy. This threat can be particularly difficult to combat because it involves the human aspect of the ecosystem. During a severe pandemic, people are anxious about their health and safety, and attackers have launched fraudulent schemes preying on COVID-19 anxieties. Coercing citizens to divulge information poses a privacy threat that could allow an attacker to identify, detect, or manipulate data.

\textbf{Data disclosure} could pose a serious privacy threat for mobile app users. In the ideal case, a server storing positive infection data does not collect any information that could potentially identify a user. In other words, even the server couldn’t learn any information about the users behind the data. However, various aspects could affect this server privacy such as the collection of metadata or the length of time data is stored, among other things. If an attacker is able to gain access to the server and thus to the data stored on the server, these factors could affect what privacy leakage occurs. Data disclosure could occur through any of the nodes where data either stays at rest or in flow.

\textbf{Eavesdropping} can be passive or active but always refers to an attacker observing communication for a malicious purpose. One scenario that could take place in the context of ENS is for an attacker to place Bluetooth devices that observe and collect transmitted contact numbers. With the knowledge of a collection device’s location and the positive infection status, an attacker could detect whether an observed device is part of the positive infection set, or could identify other devices that may have gotten an exposure notification due to the positive infection dataset. Finally, with the additional data captured during eavesdropping and the datasets available to a user, capability to combine multiple datasets, an attacker could combine information and perform a linkage attack causing additional privacy leakage. Eavesdropping could be at different layers and different stages of the operation. In addition to the cases discussed for bluetooth signal eavesdropping, an attacker could also eavesdrop on an app’s interaction with the operating system and use it to identify privacy sensitive information about the user. Such examples have been explored exhaustively in the side-channel attack literature.

\textbf{Replay attacks} re-transmit legitimate data with the goal of gaining information or disrupting typical functionality. In an ENS, this could take the form of an attacker replaying any anonymous values that were observed while listening on Bluetooth channels. This could cause devices to think they had contact with a device they did not have contact with. An attack of this type could overwhelm manual contact tracing infrastructure, or more generally cause anxiety in citizens.

\textbf{Spoofing} is an adversarial action to impersonate a user in order to either disseminate fraudulent data or collect data that does not belong to the attacker. For ENS, an attacker may attempt to spoof contact events or even positive infection statuses, affecting the integrity of data. Another form of spoofing attack is to tamper with the data. 

\textbf{Tampering} with data refers to data being changed, either to another seemingly valid value or to a meaningless value that may compromise the usefulness of a dataset. Because many ENS mobile apps are decentralized, meaning that data is processed on the devices themselves, tampering may be a bigger threat on application servers that may store data for a specific purpose related to the app functionality.

\subsection{Privacy Threat Model}
There are multiple ways to represent privacy threat models and we choose to use a simplified tabular representation. All of the threat types described below are reproduced using various tools such as mitmproxy \cite{mitmproxy} and GAEN Explorer \cite{billpugh} or using experiments done by other researchers \cite{baumgrtner2020mind, leith, gvili}. Also, we have not included any security or other adversarial attacks, such as Drain Attacks as they don’t directly impact privacy of the data subjects.

D - Direct Identifier   I - Indirect Identifier
\begin{center}
\begin{longtable}{|p{0.2\textwidth}|p{0.7\textwidth}|}
 \hline
 \multicolumn{2}{|l|}{\textbf{Attacker Capability - Coercion Attacks}} \tabularnewline 
 \hline
 \textbf{CA001} \cellcolor{Gray}
 &
Infection status or potential exposure of the person revealed in mobile apps \cellcolor{Gray}
 \\ 
 \hline
 Attacker Goals
 &
Stored data compromise, identification
 \\
 \hline
 Attackers
 &
Authorities, organizations, employers, assailants
 \\
 \hline
 Privacy Identifiers
 &
Infection status (D), exposure details (I)
 \\
 \hline
 Attack Details
 &
Attackers may coerce the user to show, in the mobile app, the infection status or potential exposure to infection. 
 \\
 \hline
 \textbf{CA002} \cellcolor{Gray}
 &
 Suitable ways of withdrawing consent are not built into mobile apps\cellcolor{Gray}
 \\
 \hline
 Attacker Goals
 &
Secondary use
 \\
 \hline
 Attackers
 &
Authorities, organizations
 \\
 \hline
 Privacy Identifiers
 &
Infection status (D), exposure details (I)
 \\
 \hline
 Attack Details
 &
Consent is used as a legal basis for data processing and the attackers can make the design of the mobile app complicated for the users trying to withdraw their consent. 
 \\
 \hline
 \textbf{CA003} \cellcolor{Gray}
 &
A law-enforcement attacker can compromise privacy using a subpoena \cellcolor{Gray}
 \\
 \hline
 Attacker Goals
 &
Stored data compromise, identification
  \\
 \hline
 Attackers
 &
Authorities, hackers, assailants
  \\
 \hline
 
Privacy Identifiers
&
Infection status (D), exposure details (I)
  \\
 \hline
 Attack Details
 &
Authorities can force the users to disclose their private information stored in the mobile app using a legal mechanism such as a subpoena. The attacker then can learn the social interactions of the user of the mobile app.
 \\
 \hline
 \textbf{CA004} \cellcolor{Gray}
 &
Users are not given sufficient information about the app \cellcolor{Gray}
  \\
 \hline
 Attacker Goals
 &
Stored data compromise, identification
  \\
 \hline
 Attackers
 &
Authorities, hackers, assailants
  \\
 \hline
 Privacy Identifiers
 &
Infection status (D), exposure details (I), and other personal information
   \\
 \hline
 Attack Details
 &
Attackers can use misinformation to force users to upload private information to a server and consequently reveal user information.
   \\
 \hline
\end{longtable}
\end{center}

\begin{center}
\begin{longtable}{|p{0.2\textwidth}|p{0.7\textwidth}|}
 \hline
 \multicolumn{2}{|l|}{\textbf{Attacker Capability - Data Disclosure}} \tabularnewline
 \hline
 \textbf{DD001} \cellcolor{Gray}
 &
Users lose control of their mobile device, allowing the people to see personal health data \cellcolor{Gray}
 \\ 
 \hline
 Attacker Goals
 &
Stored data compromise, identification
 \\
 \hline
Attackers
&
Hackers, organizers, assailants
 \\
 \hline
 Privacy Identifiers
 &
Infection status (D), exposure details (I)
  \\
 \hline
 Attack Details
 &
Users lose control of their phone, allowing people to see all the data including infection status, date of infection, and other personal data.
  \\
 \hline
 \textbf{DD002} \cellcolor{Gray}
 &
Privacy leaks while using third-party tools and technologies \cellcolor{Gray}
   \\
 \hline
 Attacker Goals
 &
Stored data compromise, identification, linkability
    \\
 \hline
 Attackers
 &
Authorities, hackers, assailants
     \\
 \hline
 Privacy Identifiers
 &
Infection status (D), exposure details (I), user details (D)
     \\
 \hline
 Attack Details
 &
Third-party tools, used in conjunction with mobile apps or web servers for various purposes like troubleshooting, are likely to disclose private data. For example, when the “Usage and diagnostics” option in Google Play Services is enabled (which it is by default), then telemetry data on GAEN operation is shared with Google.
      \\
 \hline
 \textbf{DD003} \cellcolor{Gray}
 &
Indefinite storage of data and possible later linkage with other
personal data \cellcolor{Gray}
       \\
 \hline
 Attacker Goals
 &
Stored data compromise, identification, linkability
       \\
 \hline
 Attackers
 &
Authorities, hackers, assailants
      \\
 \hline
 Privacy Identifiers
 &
Infection status (D), exposure details (I), user details (D)
   \\
 \hline
Attack Details
&
Authorities, either deliberately or through misconfigured implementations, retain the data permanently. It would be possible to link the data retroactively with other data to carry out de-anonymization attacks. 
   \\
 \hline
\end{longtable}
\end{center}

\begin{center}
\begin{longtable}{|p{0.2\textwidth}|p{0.7\textwidth}|}
 \hline
 \multicolumn{2}{|l|}{\textbf{Attacker Capability - Eavesdropping}} \tabularnewline
 \hline
 \textbf{EV001} \cellcolor{Gray}
 &
 Identification of users based on communication data \cellcolor{Gray}
 \\ 
 \hline
Attacker Goals
&
Identification, surveillance, linkability
 \\
 \hline
Attackers
&
Authorities, organizations, hackers
 \\
 \hline
 Privacy Identifiers
 &
TEK (D), IP address (I), Device information (I)
  \\
 \hline
 Attack Details
 &
When a user chooses to share TEK with the designated health authorities, the operators can identify personal details of the users by means of communication metadata such as IP address. This results in re-identification of the users and associated healthcare data like TEK.
   \\
 \hline
 \textbf{EV002} \cellcolor{Gray}
 &
Profiling user movement using daily RPID generated \cellcolor{Gray}
    \\
 \hline
 Attacker Goals
 &
Identification, surveillance, linkability
  \\
 \hline
Attackers
&
Authorities, organizations, hackers
  \\
 \hline
 Privacy Identifiers
 &
TEK (D), IP address (I), device information (I), location information (I)
   \\
 \hline
 Attack Details
 &
Attackers can build movement profiles of the users using RPIDs captured using Bluetooth LE sniffers and capture RPIDs and TEKs downloaded. Using the movement profiles, especially public locations like train stations or office locations, user de-anonymization is possible.
   \\
 \hline
 \textbf{EV003} \cellcolor{Gray}
 &
Profiling user movement using daily RPID generated \cellcolor{Gray}
    \\
 \hline
 Attacker Goals
 &
Identification, surveillance, linkability
  \\
 \hline
Attackers
&
Authorities, organizations, hackers
  \\
 \hline
 Privacy Identifiers
 &
TEK (D), IP address (I), device information (I), location information (I)
   \\
 \hline
 Attack Details
 &
Attackers can build movement profiles of the users using RPIDs captured using Bluetooth LE sniffers and capture RPIDs and TEKs downloaded. Using the movement profiles, especially public locations like train stations or office locations, user de-anonymization is possible.
     \\
 \hline
 \textbf{EV004} \cellcolor{Gray}
 &
Tracing using Bluetooth interface’s MAC address \cellcolor{Gray}
     \\
 \hline
 Attacker Goals
 &
Identification, surveillance, linkability
  \\
 \hline
Attackers
&
Authorities, organizations, hackers
     \\
 \hline
 Privacy Identifiers
 &
IP address (I), device information (I), location information (I)
     \\
 \hline
 Attack Details
 &
Smartphones with Bluetooth enabled can be traced with the help of the Bluetooth interface’s MAC address. An attacker could use this technique to trace users at specific locations. 
      \\
 \hline
\end{longtable}
\end{center}

\begin{center}
\begin{longtable}{|p{0.2\textwidth}|p{0.7\textwidth}|}
 \hline
\multicolumn{2}{|l|}{\textbf{Attacker Capability - Spoofing, Tracing, and Replay Attacks}} \tabularnewline
 \hline
\textbf{ST001} \cellcolor{Gray}
&
BLE range extensions can generate false positives \cellcolor{Gray}
 \\ 
 \hline
Attacker Goals
&
Integrity compromise
 \\
 \hline
 Attackers
 &
Hackers, authorities, organizations, employers, assailants
 \\
 \hline
 Privacy Identifiers
&
Infection status (D), location (I)
 \\
 \hline
Attack Details
&
To cause false alarms, an attacker can place BLE range extenders to spoof that a device is “nearby” when it may not be. To complete the attack, the attacker must ensure that the interactions between the attacker's device and other devices are logged as a contact event. 
\\
\hline
\textbf{ST002} \cellcolor{Gray}
&
Generate false alarms though active relays \cellcolor{Gray}
\\
\hline
Attacker Goals
&
Integrity compromise
 \\
 \hline
 Attackers
 &
Hackers, authorities, organizations, employers, assailants
 \\
 \hline
Privacy Identifiers
&
Infection status (D), location (I)
\\
\hline
Attack Details
&
An attacker can generate a false alarm by real-time relay attacks that lead to the users falsely being alerted that they may have had a positive exposure. One attacker tactic could be via the relay of Bluetooth signals from the people tested at testing centers to other phones in the proximity.
\\
\hline
\textbf{ST003} \cellcolor{Gray}
&
Generate wormhole attacks \cellcolor{Gray}
\\
\hline
Attacker Goals
&
Integrity compromise, misattribution, exclusion
\\
\hline
 Attackers
 &
Hackers, authorities, organizations, employers, assailants
 \\
 \hline
 Privacy Identifiers
&
Infection status (D), location (I)
 \\
 \hline
 Attack Details
 &
Using commercially available Bluetooth LE sniffers, attackers can collect RPIDs and pass them onto more distant locations without being noticed. This compromises the contact tracing system as a whole by falsely duplicating information about the presence of infected persons in many locations. 
 \\
 \hline
 \textbf{ST004} \cellcolor{Gray}
 &
A verification code used to upload information of another app \cellcolor{Gray}
  \\
 \hline
 Attacker Goals
 &
Stored data compromise, identification
   \\
 \hline
 Attackers
 &
Authorities, service provider, employers, hackers
  \\
 \hline
 Privacy Identifiers
 &
Infection status (D)
   \\
 \hline
 Attack Details
 &
An attacker can use a verification code, generated for an infected person to upload their temporary encryption key (TEK), in another app to upload their information, resulting in compromise of the overall system. 
  \\
 \hline
\end{longtable}
\end{center}
\subsection{Threat Mitigation Strategy}

We propose Privacy by Design methodology \cite{cavoukian}, developed by  Anne Cavoukian, Ontario’s Data Protection Commissioner, because of its applicability in mitigating conventional and non-stereotypical threats. Also, we didn’t provide a prescriptive threat mitigation strategy because the mitigation strategy can be applied at various levels, ranging from legal frameworks to technical solutions like differential privacy for data stored in the servers.  

The following table lists all mitigation strategies and privacy attacks the respective mitigation strategy addresses:

\begin{center}
\begin{longtable}{|p{0.24\textwidth}|p{0.4\textwidth}|p{0.29\textwidth}|}
 \hline
\textbf{Mitigation Strategies} \cellcolor{Gray}
 & \textbf{Details} \cellcolor{Gray}
 & \textbf{Privacy Attacks Addressed} \cellcolor{Gray}
 \\ 
 \hline
 \multirow{7}{9em}{Applying a Privacy-by-Design approach for developing ENS solution components (mobile app, backend server, etc.)}
 &
 Data Minimization: Identify and use only required attributes. Remove / destroy any data, whether direct identifier or indirect identifiers, when they are no longer required. The cryptographic tokens shared by the apps should be valid for the shortest interval and should be continuously replaced with new tokens, so that an app cannot be tracked using the tokens it shares. 
 &
 DD02, DD003
  \\ \cline{2-3}
 &
 Data Separation: Isolate and distribute processing of personal data. Processing separation to be applied at different components of DCT including separation of verification activities from exposure notifications \cite{billpugh}. Further levels of privacy can be attained by using methods like multi-party computation or secure enclaves in order to ensure privacy of data during computation.
&
DD002
  \\ \cline{2-3}

 &
 
Data Abstraction and Perturbation: Aggregate and add noise to the data at source. For example, it is recommended that data is abstracted or perturbed before the data is uploaded from mobile app to the servers. Usage of standard metrics and mechanisms for perturbation like differential privacy should be used.
&
DD002, DD003
  \\ \cline{2-3}

 &
 Inform: Inform users and data subjects about the processing of their personal data in a timely and detailed manner.
 &
 CA004
  \\ \cline{2-3}

 &
 Control: Put users and data subjects in control over processing of their personal data and provide easy-to-use interfaces to opt-out from data sharing.
 &
 CA001*, CA002*, CA003*, DD001*
  \\ \cline{2-3}

 &
 Visibility: Provide visibility into how the personal data is processed and what privacy-enhancing techniques used.
 &
 
  \\ \cline{2-3}

 &
 Regulation: Impose strict data processing and usage regulations at the institutional level and establish role based access control within an organization.
 &
 
    \\ 
 \hline
 Securing communication between various components
 &
 Enabling SSL certificates for all data-in-transit communication.
 &
 EV001
  \\ \cline{2-3}

 &
 Removing HTTP headers like ‘Authorization’ reduces linkability issues. 
 &
 EV001
  \\ \cline{2-3}

 &
 HMAC enhancements to mitigate relay and replay attacks. \cite{cavoukian}
 &
 EV002*,EV003*,ST002*,\newline ST001*
   \\ 
 \hline
 Use Reference Architectures for ENS
 &
 Follow proven reference architecture recommendations \cite{gvili} like separation of processing, authentication, etc. 
 &
 EV002*, EV003*, ST004
  \\ \cline{2-3}

 &
 Keep mobile apps and server components current with latest APIs, patches, and OS upgrades.
 &
 EV004
  \\ \cline{2-3}
\hline
 
 & 
 Perform end-to-end security and vulnerability assessment of the DCT components.
 &
 ST001*, ST002*, ST004*
   \\ 
 \hline
\end{longtable}
*Minimizes the attack and does not fully mitigate the attack

\end{center}

\section{Conclusion}
Although there are existing threat models which aim to address privacy challenges, the unique ecosystem within exposure notification systems requires a more comprehensive approach. To this end, we introduce an enhanced framework for Target State Privacy Threat Modeling, which simplifies terminology, minimizes overlooked threats, and ensures consistency across instances of the threat model.

Further, we used our newly enhanced framework to build a Target State Privacy Threat Model for ENS. This model reflects the comprehensive privacy concerns faced by exposure notification systems and defines specific privacy threats based on attacker capabilities, including what privacy identifiers are affected and how the attack might take place.

Finally, we provide mitigation strategies that should be taken to address the threats identified in the ENS Target State Privacy Threat Model.

\section*{Glossary}

    \begin{longtable}{|p{0.18\textwidth}|p{0.32\textwidth}|p{0.48\textwidth}|}
    \hline
    \textbf{Term/Acronym} \cellcolor{Gray}
    & \textbf{Description} \cellcolor{Gray}
    & \textbf{Reference} \cellcolor{Gray}
    \\
    \hline
    Privacy Identifiers
    & Any data, element, or information that can be used, directly or as a proxy,  to identify a person or group of persons.
    &
    
    \\
    \hline
    LINDDUN
    &
    Systematic elicitation and mitigation of privacy threats in software systems
    &
    \url{https://www.linddun.org/}
    \\
    \hline
    STRIDE &
    approach to threat modeling &
    \url{https://en.wikipedia.org/wiki/Threat_model}
    \\
    \hline
    PASTA & 
    Process for Attack Simulation and Threat Analysis (PASTA) seven-step, risk-centric methodology. &
    \url{https://en.wikipedia.org/wiki/Threat_model}
    \\
    \hline
    NIST
    &
    physical sciences laboratory and a non-regulatory agency of the United States Department of Commerce.
    &
    \url{https://en.wikipedia.org/wiki/National_Institute_of_Standards_and_Technology}
    \\
    \hline
    DCT
    &
    Digital Contact Tracing
    &
    \url{https://en.wikipedia.org/wiki/Digital_contact_tracing}
    \\
    \hline
    GDPR
    &
    General Data Protection Regulation (GDPR) has been the gold standard for personal data protection
    &
    \url{https://en.wikipedia.org/wiki/General_Data_Protection_Regulation}
    \\
    \hline
    EDPB
    &
    European Data Privacy Board
    &
    \url{https://en.wikipedia.org/wiki/European_Data_Protection_Board}
    \\
    \hline
    TAN
    &
    Acronym for "[Transaction Authentication Number]
    &
    \url{https://en.wikipedia.org/wiki/Transaction_authentication_number}
    \\
    \hline
    SMS
    &
    Short message service
    &
    \url{https://en.wikipedia.org/wiki/SMS}
    \\
    \hline
    DFD
    &
    Data Flow Diagrams
    &
    \url{https://en.wikipedia.org/wiki/Data-flow_diagram}
    \\
    \hline
    IFD
    &
    Information flow diagram
    &
    \url{https://en.wikipedia.org/wiki/Information_flow_diagram}
    \\
    \hline
    TPTM
    &
    Target Privacy Threat Modeling Framework
    &
    \url{https://insights.sei.cmu.edu/sei_blog/2018/12/threat-modeling-12-available-methods.html}
    \\
    \hline
    COVID
    &
    A respiratory illness. It is colloquially known as coronavirus.
    &
    \url{https://en.wikipedia.org/wiki/Coronavirus_disease_2019}
    \\
    \hline
    TEK
    &
    Temporary Encryption Key - used as part of the anonymous exposure notification protocols.
    &
    \url{https://en.wikipedia.org/wiki/Key_(cryptography)}
    \\
    \hline
    \end{longtable}
    \label{tab:metrics}

\bibliography{references} 
\bibliographystyle{plain}
\end{document}